# Discovery and the QuarkNet Data Portfolio

Kenneth Cecire[1]

*Department of Physics*
*University of Notre Dame*
*225 Nieuwland Science Hall*
*Notre Dame, Indiana, U.S.A.*

Student and public understanding of new discoveries in particle physics are enhanced by preparatory activities. Such activities give the user experience and context to understand a representation of the data associated with the discovery and some familiarity with the topics associated with the discovery. The QuarkNet Data Portfolio is a developing model of high-quality activities that address this need.

PRESENTED AT

DPF 2015
The Meeting of the American Physical Society
Division of Particles and Fields
Ann Arbor, Michigan
August 4–8, 2015

[1]Work supported by the National Science Foundation and the Department of Energy, Office of Science

## 1. IT IS ALREADY TOO LATE TO EXPLAIN WHEN A DISCOVERY IS ANNOUNCED.

If pre-university students and the general public are to understand a new discovery beyond a superficial level, they must be prepared ahead of time. This preparation does not need to deal so much with the specifics of the discovery—after all, that part is new to physicists as well—but with general background in the Standard Model and on how to understand evidence. Thus discovery plots, which are rich in meaning to particle physicists, may be lost to others.

The discovery of the Higgs boson is a case in point. In some ways, the public was prepared in that it was widely explained that the Higgs was an important piece of the particle puzzle that was thus far missing and that it had meaning in the question of how other fundamental particles have their masses. Yet the evidence for the discovery makes little sense to the public or to students unless they had been exposed to a program like International Masterclasses in particle physics [1].

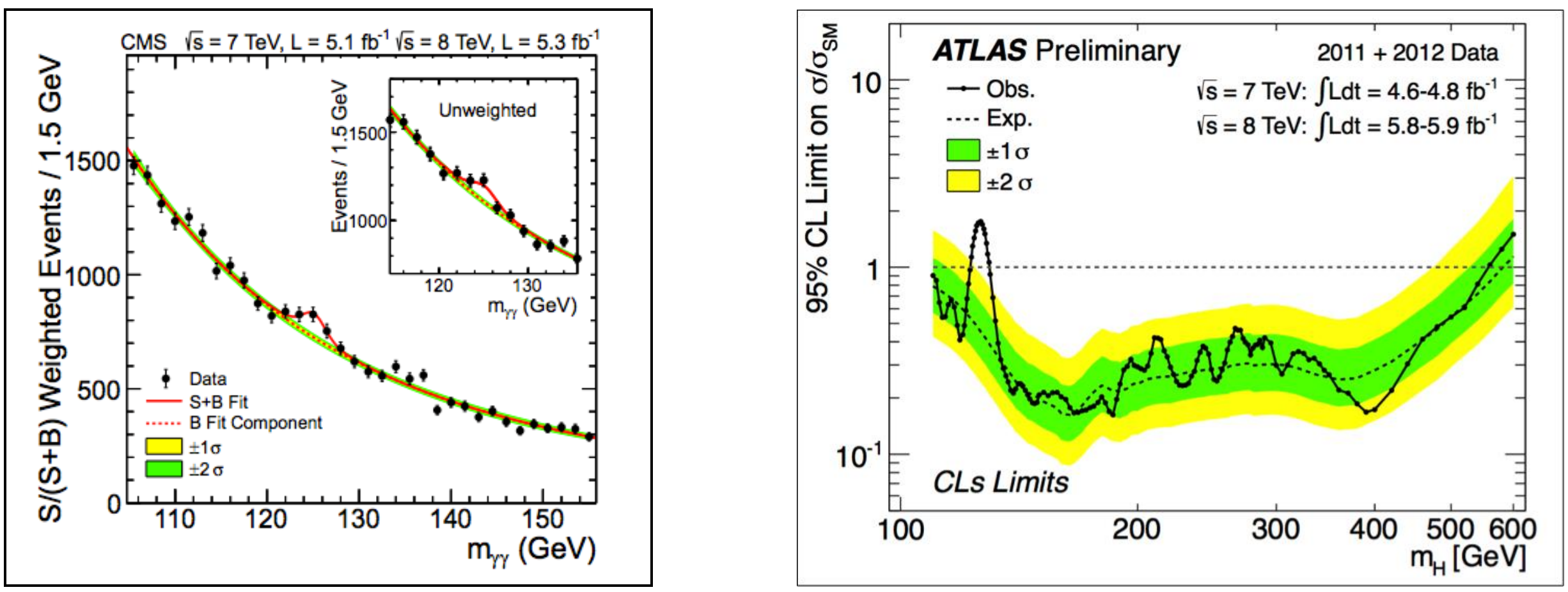

Figure 1: Two plots associated with the discovery of the Higgs boson from the CMS [2] and ATLAS [3] collaborations. The CMS plot on the left requires a great deal of explanation, which may not succeed, unless the viewer is familiar with a mass histogram. Those who are familiar with such a plot would need just as much ramp up again to understand the exclusion plot on the right.

## 2. RESOURCES ARE AVAILABLE.

### 2.1. Present General Principles Now, before the Next Discovery.

One activity can help someone to understand the next discovery. Long-term engagement can make a very large positive difference in understanding. The QuarkNet Data Portfolio [4] has a range of resources to educate students and others in how particle physicists look at and interpret data, as well as introduce how the Standard Model works. The resources are professionally vetted to meet best practices in science education and transmit key understandings in particle physics. The activities are aligned with Next Generation Science Standards and ranked into three levels of sophistication. Level 1 is introductory and deals with relatively few data points; level 2 has substantially large sets of data and requires more analysis (masterclasses are at this level); level 3 uses very large datasets, allows for longer-term study, and gives a wider range of inquiry.

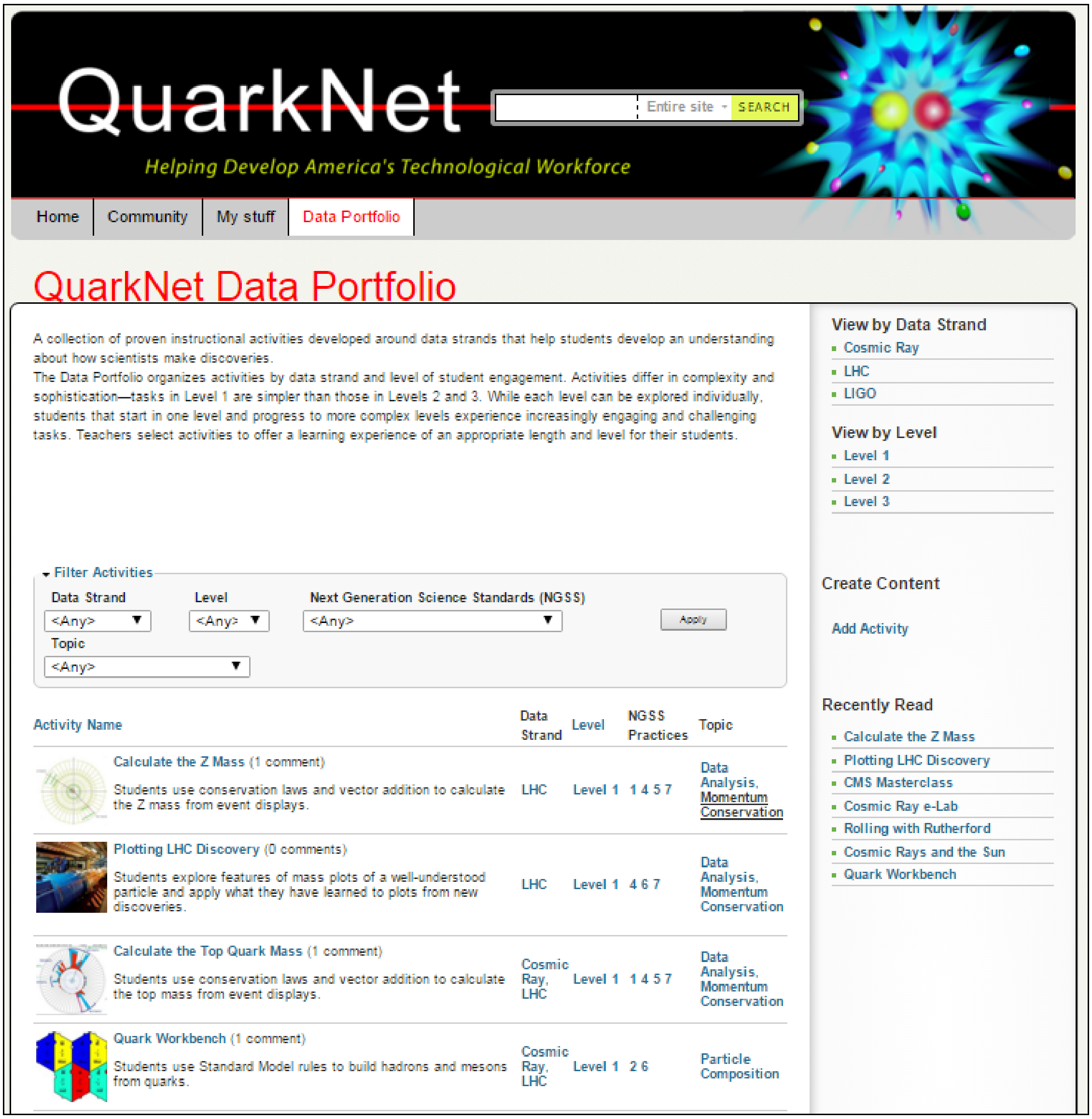

Figure 2: Front page of the QuarkNet Data Portfolio.

## 2.2. Plotting LHC Discovery

A Data Portfolio activity, which is directed specifically at the purposes of this paper, is *Plotting LHC Discovery*. Students are first given data to make a simple mass histogram of the J/Ψ meson with background events. They then distinguish the background model from the rather substantial “bump” at the mass of the J/Ψ.

**PLOTTING LHC DISCOVERY**
**INTERPRET THE LATEST FROM LHC.**

| mass (GeV) | bg model | data | data-bg |
|---|---|---|---|
| 2.7 | 7 | 7 | 0 |
| 2.8 | 7 | 7 | 0 |
| 2.9 | 7 | 8 | 1 |
| 3.0 | 6 | 17 | 11 |
| 3.1 | 6 | 62 | 56 |
| 3.2 | 6 | 13 | 7 |
| 3.3 | 6 | 6 | 0 |
| 3.4 | 5 | 5 | 0 |

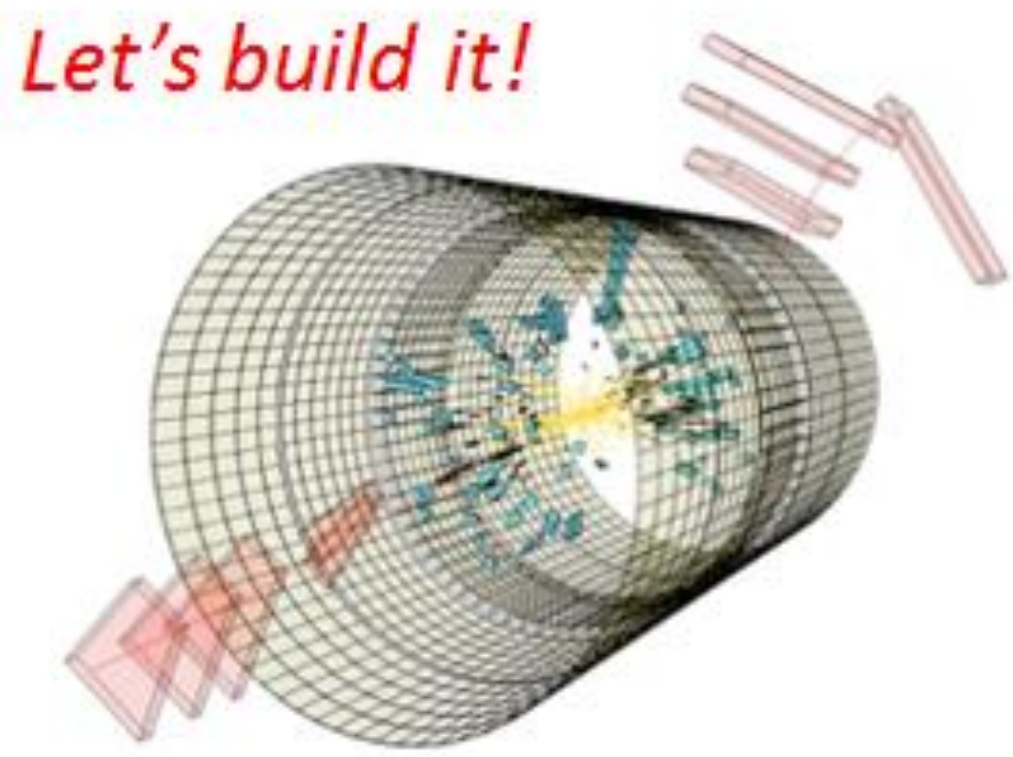

Figure 3: Simplified data table for Plotting LHC Discovery. Students assemble this data into a histogram.

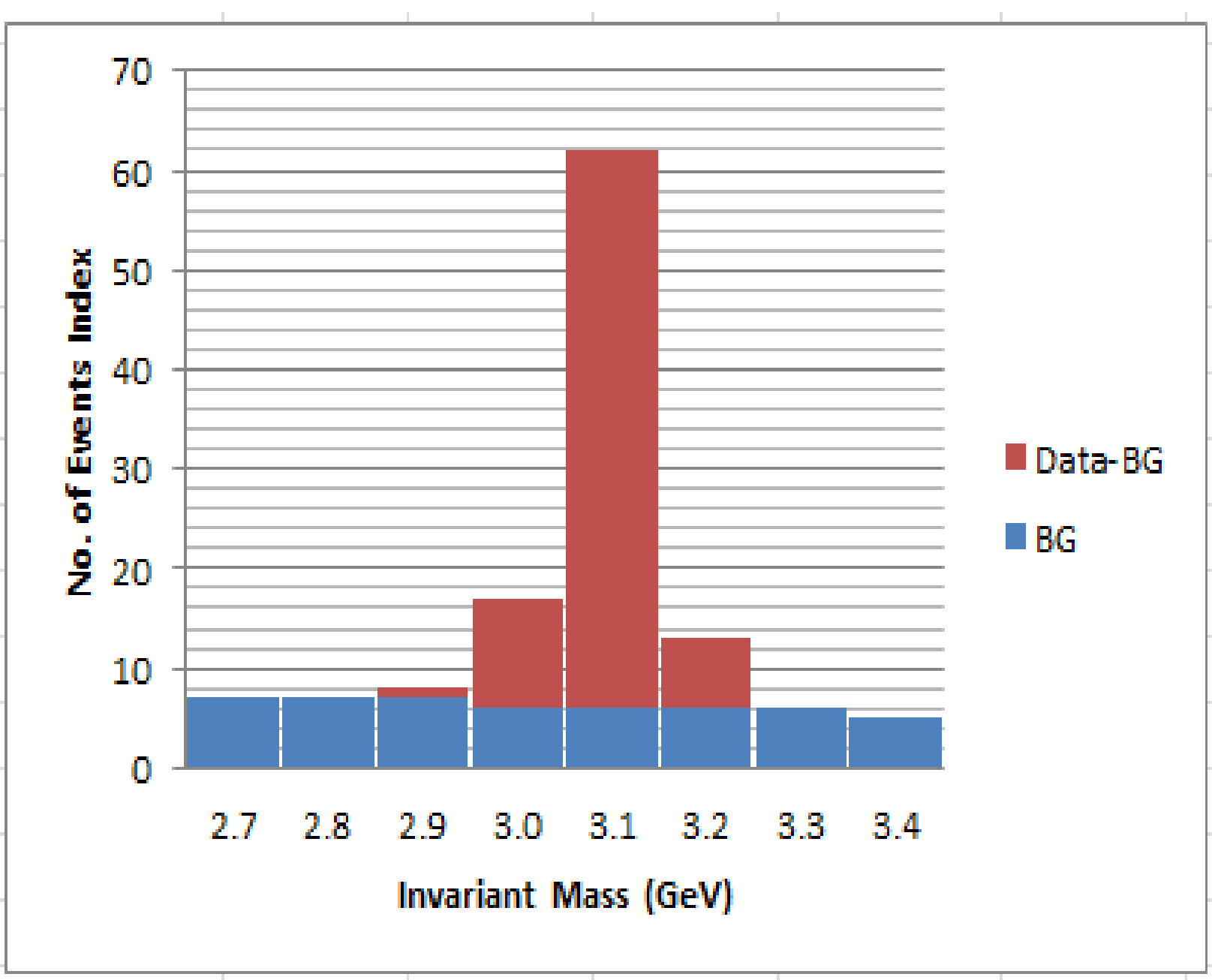

Figure 4: Mass histogram from above data. Blue correspomds to the background model while the red excess shows that there is a particle resonance at 3.1 GeV.

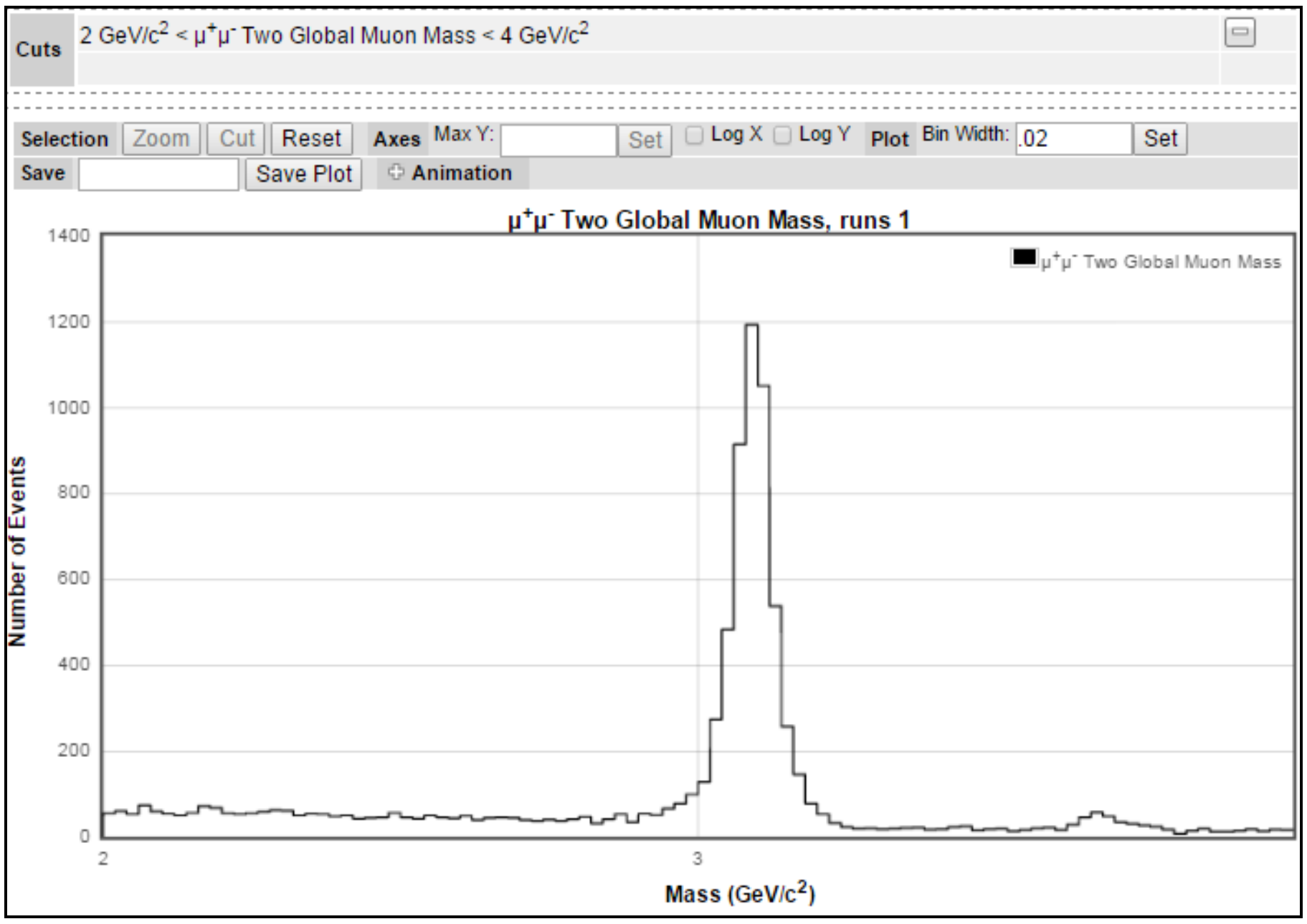

Figure 5. A J/Ψ mass histogram similar to Figure 4 but with more data, produced in the QuarkNet CMS e-Lab [5]. Signal and background are still seen. An additional smaller bump for the Ψ' meson is also just visible on the rightward portion of the plot.

Students learn, in *Plotting LHC Discovery*, to identify signal and background, to find the mass of the particle from the peak of the signal, and to qualitatively evaluate the significance of the peak from its width and its height relative to the background. They then apply this to a relatively recent discovery plot to find the same features and evaluate it. This puts them in an improved position to understand and appreciate a new discovery when it is announced.

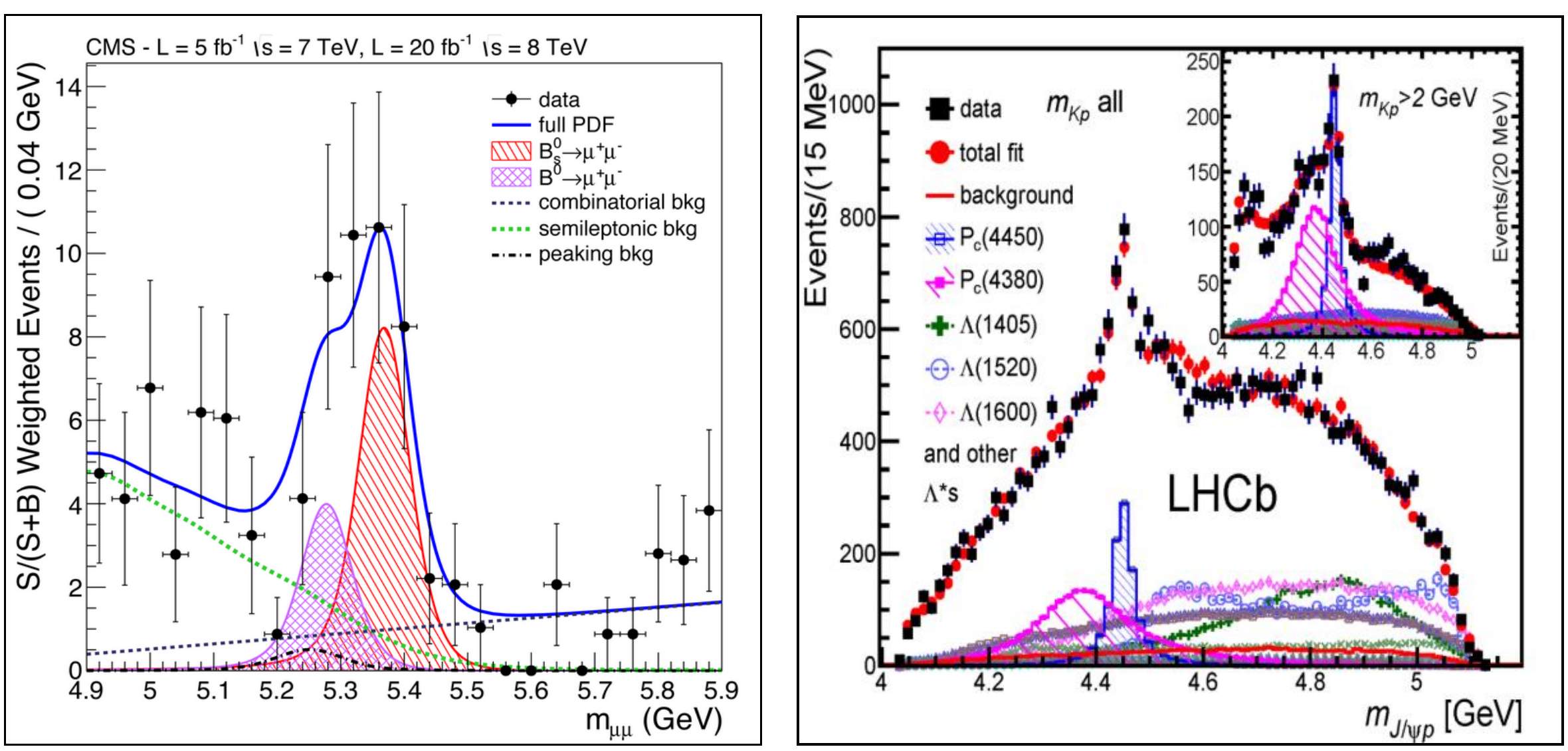

Figure 6: Two recent LHC discovery plots. The plot on the left shows rare $B_s$ and $B^0$ mesons in CMS (also seen in LHCb) [6] while the right shows a possible pentaquark signal in LHCb [7].

Actual discovery plots, such as the 2015 CMS and LHCb plots in Figure 6, are substantially more complicated than the J/Ψ signals in Figures 4 and 5, but experience with an activity like *Plotting LHC Discovery* gives students background to understand an explanation of such a plot and to be able to identify key features.

## 2.3. Masterclasses

International Masterclasses present another, complimentary way to help participants be able to understand particle physics data and prepare for discovery explanations. In International Masterclasses, students have the opportunity to analyze event displays from LHC data and use these to build statistical results, including mass histograms. Students and others who have been in an all-day masterclass have the experience of being "particle physicists for a day."

Participants who have been in International Masterclasses have learned to understand particle collision event displays and then build and analyze mass histograms from them. A discovery mass plot would therefore likely be accessible to them.

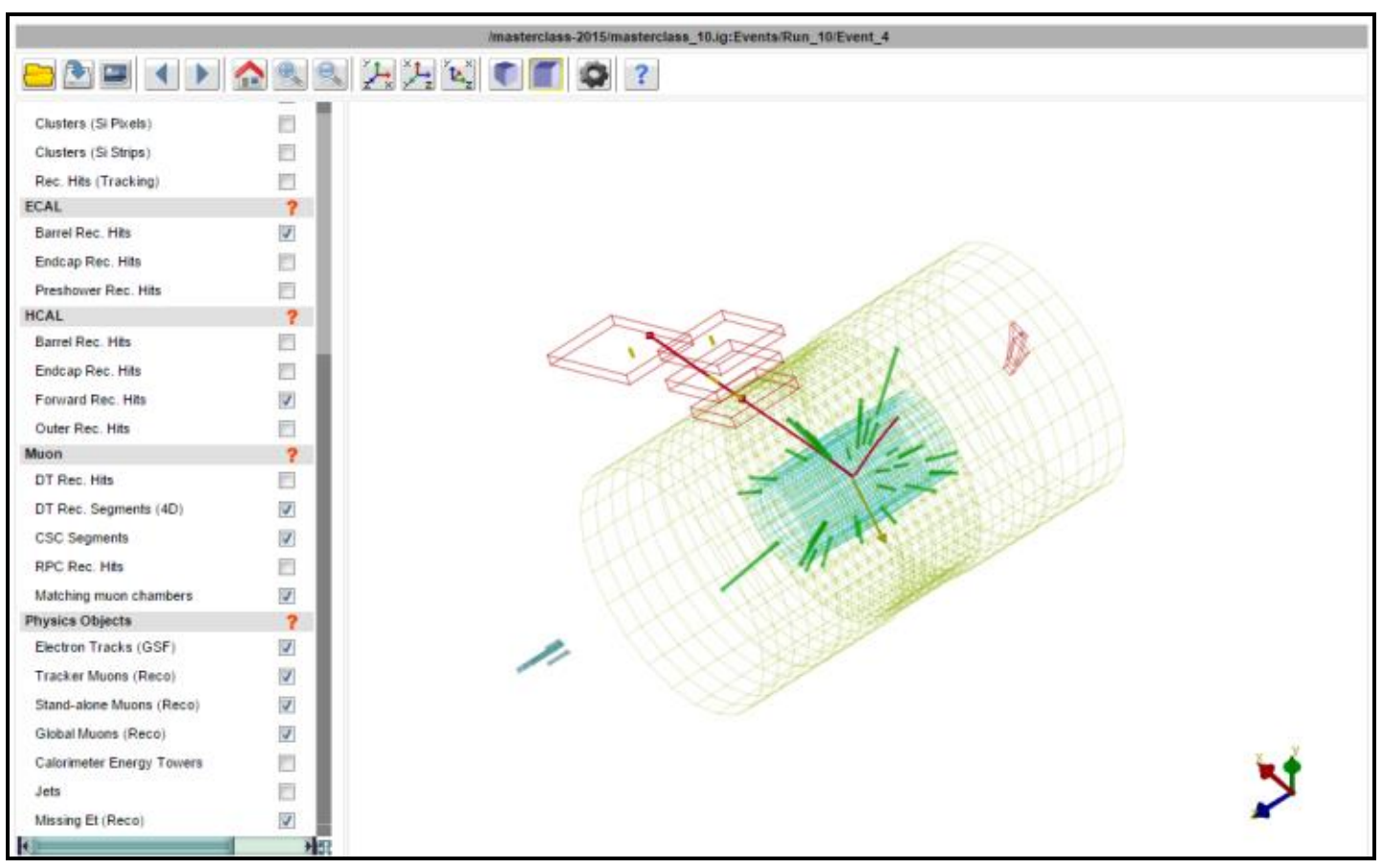

Figure 7: Event display from the CMS WZH-path masterclass.

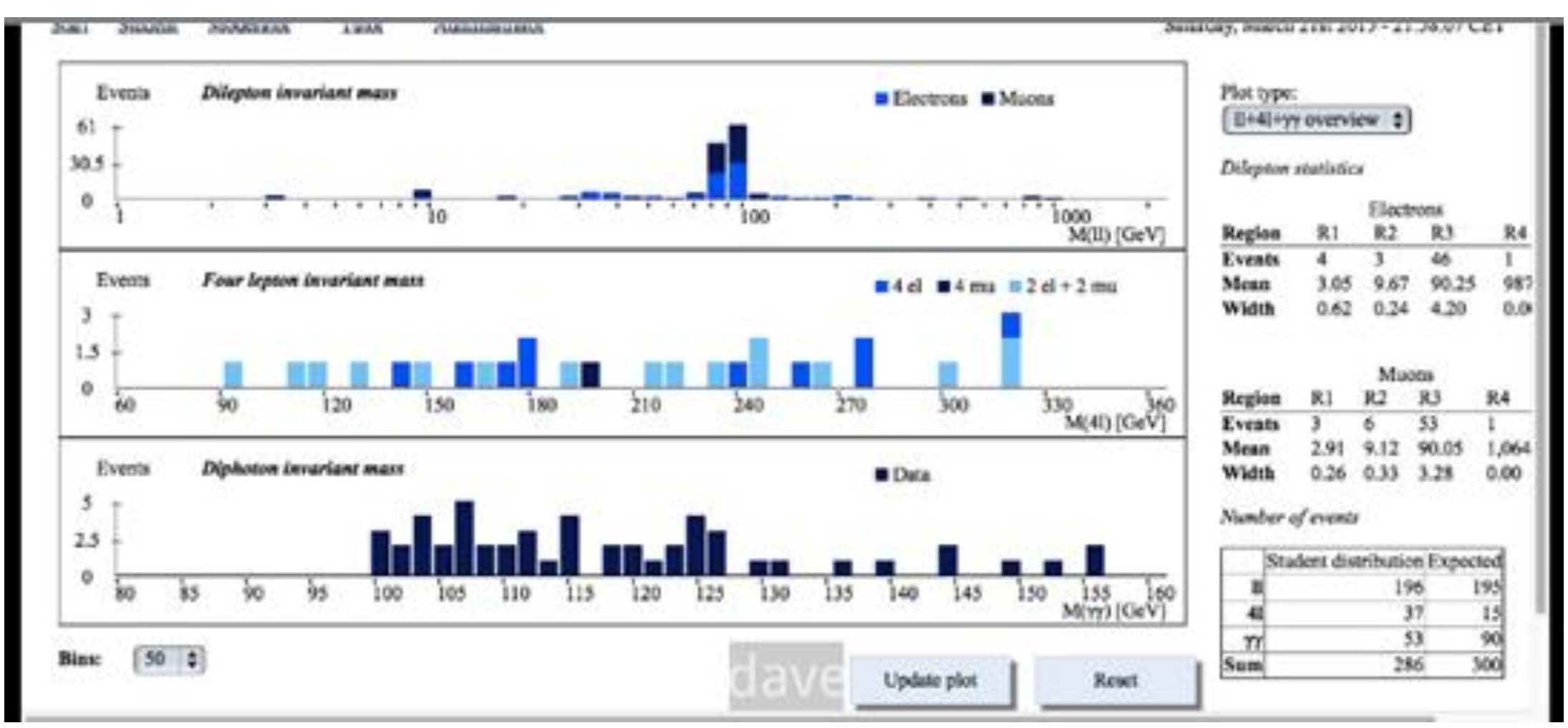

Figure 8: Mass histograms – dilepton, four lepton, and diphoton, respectively, from the ATLAS Z-path masterclass.

## Acknowledgments

The author wishes to thank colleagues in the QuarkNet program and in International Masterclasses for the material that makes up this paper. He especially wishes to thank, posthumously, QuarkNet Coordinator Tom Jordan, who passed away in Spring 2015, without whose work on the QuarkNet Data Portfolio this paper would have not been possible.

Work is supported by the National Science Foundation QuarkNet grant, award #1219444.